\documentclass[conference]{IEEEtran}
\IEEEoverridecommandlockouts
\usepackage{cite}
\usepackage{amsmath,amssymb,amsfonts}
\usepackage{algorithmic}
\usepackage{graphicx}
\usepackage{textcomp}
\usepackage[caption=false, font=footnotesize, labelformat=empty]{subfig}

\def\BibTeX{{\rm B\kern-.05em{\sc i\kern-.025em b}\kern-.08em
		T\kern-.1667em\lower.7ex\hbox{E}\kern-.125emX}}

\renewcommand{\baselinestretch}{0.905}

\begin{document}
	
	\title{\huge Coexistence of Continuous Variable Quantum Key Distribution and 7$\times$12.5 Gbit/s Classical Channels\vspace{-7pt}
	}
	
	\author{ \IEEEauthorblockN{Tobias A. Eriksson$^{(1)}$, Takuya Hirano$^{(2)}$,  Motoharu Ono$^{(2)}$, Mikio Fujiwara$^{(1)}$, Ryo Namiki$^{(2)}$,\\  Ken-ichiro Yoshino$^{(3)}$, Akio Tajima$^{(3)}$, Masahiro Takeoka$^{(1)}$, and Masahide Sasaki$^{(1)}$}
		\IEEEauthorblockA{\small ${(1)}$
			\textit{National Institute of Information and Communications
				Technology (NICT)},
			\textit{4-2-1 Nukui-kitamachi, Koganei, Tokyo 184-8795, Japan.}  \\
			${(2)}$ \textit{Department of Physics, Gakushuin University, 1-5-1 Mejiro,Toshima-ku,Tokyo, 171-8588, Japan.}\\
			${(3)}$ \textit{IoT Devices Research Labs., NEC Corporation, 1753 Shimonumabe, Nakahara-ku, Kawasaki 211-8666, Japan.}\\
			e-mail: eriksson@nict.go.jp }\vspace{-10pt}
	}
	\maketitle

\begin{abstract}
We study coexistence of CV-QKD and 7 classical 12.5 Gbit/s on-off keying channels in WDM transmission over the C-band. We demonstrate key generation with a distilled secret key rate between 20 to 50 kbit/s in experiments running continuously over 24 hours.
\end{abstract}

\vspace{-5pt}
\section{Introduction}\vspace{-0pt}
Quantum key distribution (QKD) is the first quantum technology to find commercial application and it is the only known solution to the problem of sharing a random key between two distance parties with proven security against any possible eavesdropping attack. QKD relies on quantum mechanical properties of light using either single photons (discrete variable (DV)) or weak coherent pulses (continuous variable (CV)). With QKD, information theoretic security can be guaranteed which is in stark contrast to current generation of encryption techniques that relies on computationally hard problems. 

The first proposed QKD scheme is the BB84 protocol \cite{BB84}, which utilizes the polarization states of single photons to convey a secret key between two parties. Today, many different protocols exist for DV-QKD with the main common factor being that they rely on single photon detection \cite{DiamantiPratical}. Several different DV-QKD techniques have been demonstrated over installed fiber \cite{SasakiTokyoQKD} as well as over free-space links \cite{SchmittManderbach}. CV-QKD on the other hand, can to a large extent be implemented with commercially available telecommunication components. Further, it is also compatible with photonic integration techniques, making CV-QKD a promising technology for future QKD systems. The potential of CV-QKD has been demonstrated in several experimental investigations \cite{Lodewyck, JouguetNature, Huang, Nakazawa}.

One key challenge for QKD systems is to manage integration into the current network topology, to avoid having to develop a separate QKD network \cite{DiamantiPratical}. This means that CV-QKD has to be able to coexist with classical wavelength devision multiplexed signals in fiber optical links. Here, CV-QKD has a big advantage due to the mode selection capability when using detection techniques based on homodyne or heterodyne receivers with a local oscillator. Coexistence of DV-QKD at 1300~nm and 64-QAM channels in the C-band have been reported in \cite{Wang}. In \cite{Kumar}, the influence of continuous wave lasers and CV-QKD channels multiplexed in the C-band is experimentally investigated. Further, the excess noise in a 20 channel wavelength division multiplexing (WDM) system has been investigated in \cite{Karinou}.

In this paper, we demonstrate WDM co-propagation in the C-band of a 10 MHz repetition rate CV-QKD system together with 7 neighboring classical on-off keying data channels at 12.5 Gbit/s each within the C-band. We show stable key generation over 10~km of fiber for 24~hours. 

\vspace{-5pt}
\section{Experimental Setup}
\begin{figure}[]
	\centering
	\includegraphics[width=1\columnwidth]{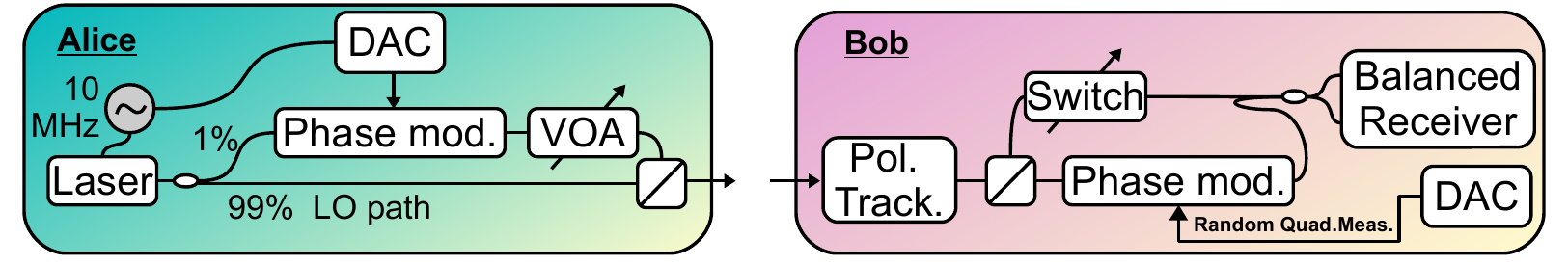}	\vspace{-20pt}%
	\caption{Rough overview of the optical part of the CV-QKD system. The transmitter laser is pulsed at 10~MHz, 99\% of the intensity is coupled to one polarization to be used as a local oscillator (LO) while the remaining 1\% is coupled to the state preparation stage where four states are modulated by a phase modulator driven by a digital analogue converter (DAC). At Bob's receiver active polarization tracking is performed before the quantum signal and the LO is split using a polarization beam splitter.  The LO path is directed to a phase modulator which randomly choose quadrature to measure in the homodyne detector. The signal path is directed to a optical switch which can block the incoming CV-QKD signal to characterize the shot noise of the receiver. The signal and LO are mixed before homodyne (single quadrature) detection is performed with a balanced receiver module.}
	\label{fig:CVQKDsetup}\vspace{-5pt}%
\end{figure}

\vspace{-0pt}
The outline of the CV-QKD system is shown in Fig.~\ref{fig:CVQKDsetup}(a), for details see the figure text. The system applies the four state CV-QKD protocol together with reverse reconciliation. Not shown in the figure is the synchronization signal that is transmitted at 1300~nm that adjusts the sampling instance of Bob's analog to digital converter. For more detailed information on the CV-QKD system and  the key distillation process, please see \cite{Hirano}.

\begin{figure}[]
	\vspace{-5pt}%
	\centering
	\includegraphics[width=1\columnwidth]{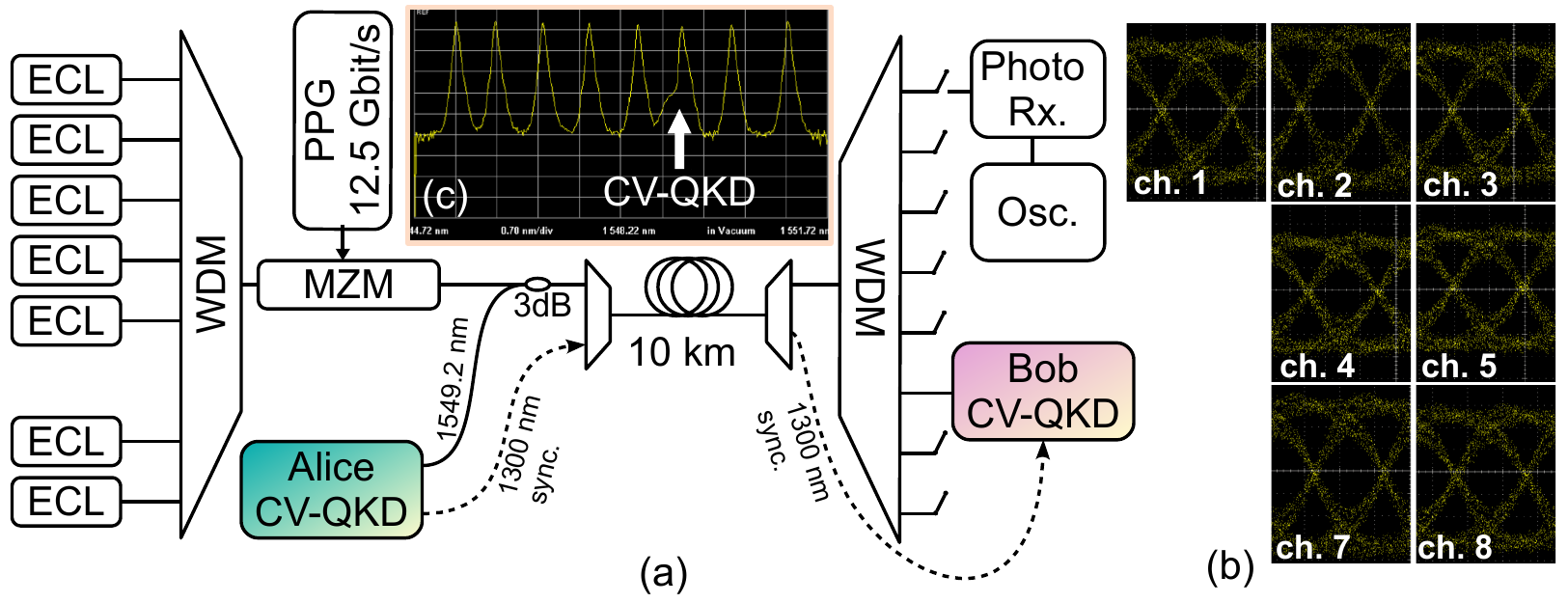}
	\vspace{-20pt}%
	\caption{WDM transmission experiment setup. 7 ECLs placed on an 100~GHz frequency grid. The CV-QKD system uses the 6th channel of the WDM setup. }
	\label{fig:expSetup}\vspace{-15pt}%
\end{figure}

\begin{figure*}[t]
	\addtocounter{figure}{2}
	\centering
	\includegraphics[width=1\textwidth]{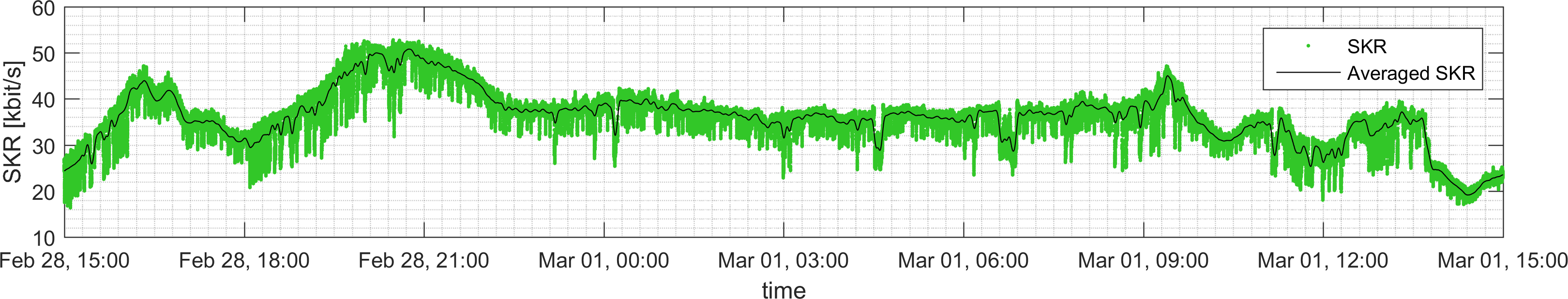}
	\vspace{-17pt}
	\caption{Distilled secret key rate over 24 hours.  }
	\label{fig:24hours}\vspace{-5pt}%
\end{figure*}

The overview of the experimental setup is shown Fig.~\ref{fig:expSetup}. The WDM system is using a channel spacing of 100~GHz. We use 7 external cavity lasers (ECLs) that are modulated with 12.5~Gbit/s on-off keying data using a pseudo random bit sequence of length $2^{15}-1$, constituting the channels 1-5 and 7-8 of the WDM system. The CV-QKD signal is using a wavelength of 1549.2~nm and is transmitted in the 6th band of the WDM system. The transmitted spectra is shown in Fig.~\ref{fig:expSetup}(c). The difference in launch power of the channels are less than 1.5~dB and is on average -4.5 dBm.  The data channels and the CV-QKD channel are coupled together using a 3~dB coupler. Further, the CV-QKD system transmits a synchronization pulse for the clock signal at 1300~nm. This signal is coupled together with the C-band channels using a WDM coupler.

\begin{figure}[]
	\addtocounter{figure}{-3}
	\vspace{-20pt}
	\centering
	\includegraphics[width=0.99\columnwidth]{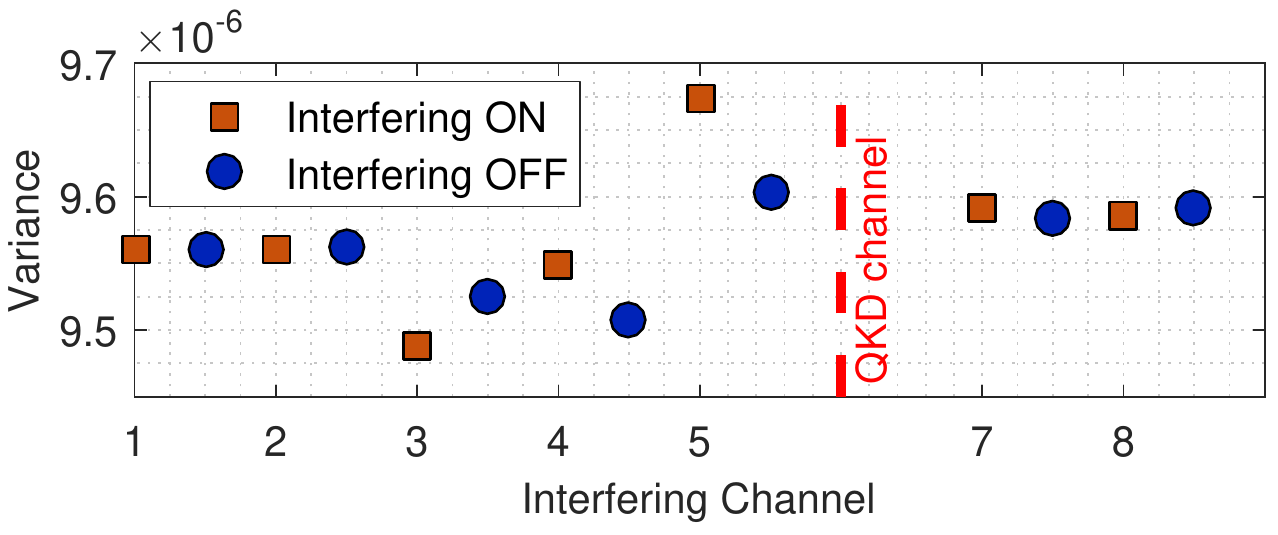}\vspace{-8pt}
	\caption{Variance as a function of interfering channel where one modulated channel is turned on at the time. }
	\label{fig:variance}\vspace{-5pt}%
\end{figure}

The signals are transmitted over 10~km of conventional single mode fiber. At the receiver, the CV-QKD sync. pulse at 1300~nm is first demultiplexed using a WDM coupler and sent to the CV-QKD receiver. Further, the data channels and the CV-QKD signal are demultiplexed using an 8 channel WDM demultiplexer. The data channel can be switched to a 10~GHz photodetector and digitized by an 8~GHz bandwidth oscilloscope. The CV-QKD wavelength is always directed to the CV-QKD receiver where the secret key rate (SKR) is evaluated.

\begin{figure}[]
	\centering
	\includegraphics[width=0.99\columnwidth]{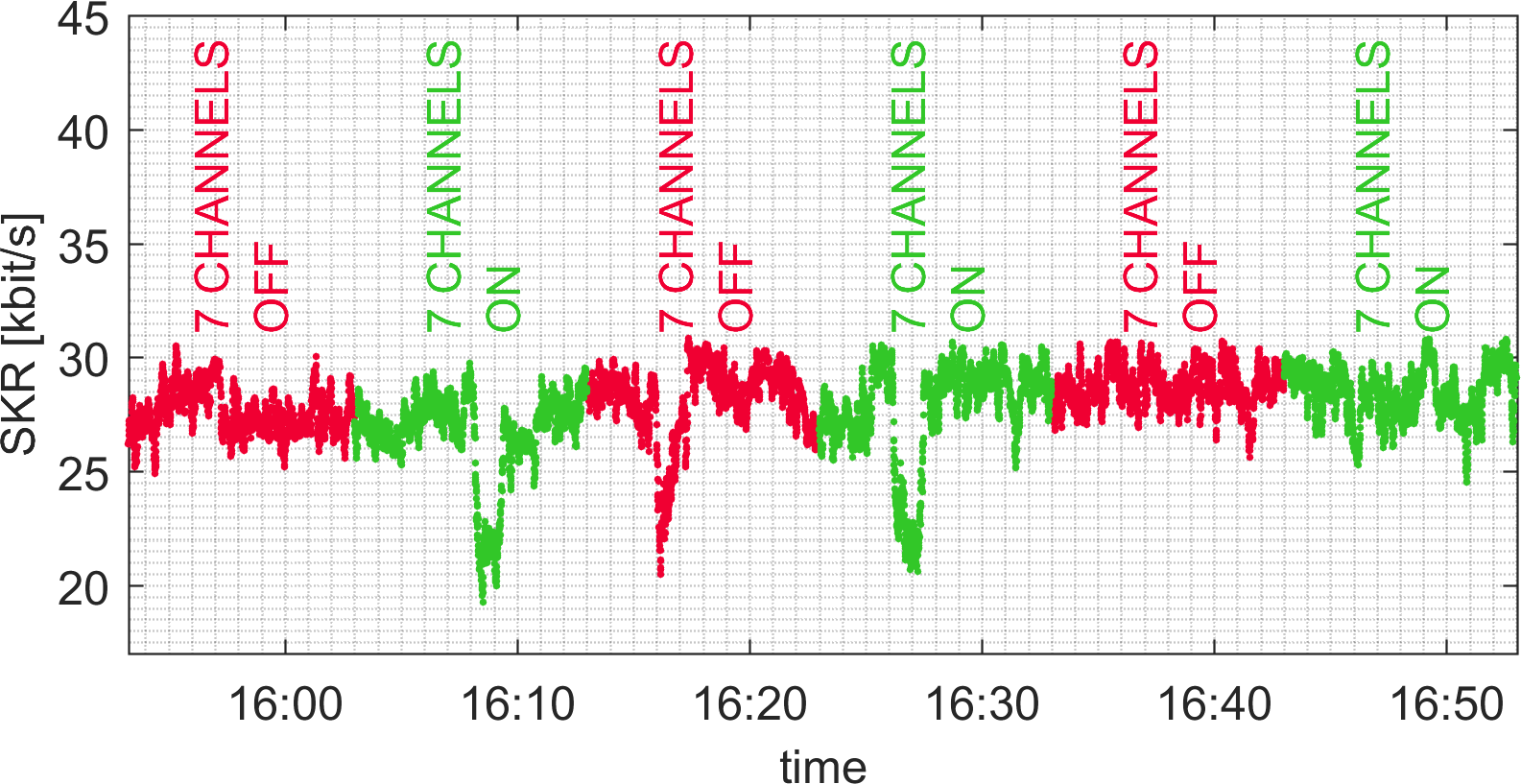}\vspace{-8pt}
	\caption{Secret key rate as a function of time when the 7 co-propagating classical channels are turned on and off in intervals of 10 minutes. }
	\label{fig:onfoff}\vspace{-15pt}%
\end{figure}

\vspace{-5pt}
\section{Results}\vspace{-5pt}
Our initial experiment deviates slightly from the experimental setup in Fig.~\ref{fig:expSetup}. In this case we are modulating a single wavelength before the WDM multiplexer, also Alice's CV-QKD signal goes through channel 6 on the WDM multiplexer. In this experiment we measure the variance of the received signals when we turn on different co-propagating WDM channels, one at a time. The results are shown in Fig.~\ref{fig:variance}. We average the variance over data taken for at least 3 minutes. The maximum change in SKR when turning on one of the channels is 0.75\%, which is within our measurement reliability. 

These results encouraged us to perform the experiment with all 7 co-propagating channels on, using the setup in Fig.~\ref{fig:expSetup}. We first investigate the impact from turning on and off the 7 co-propagating channels. The SKR as a function of time is shown in Fig.~\ref{fig:onfoff} where turn on and off the 7 channels with intervals of 10 minutes. We do not notice any influence on the SKR from the co-propagating channels. Some drops in the SKR are observed, but these occur both when the co-propagating channels are on as well as when they are off. These jumps are attributed to the system being disturbed by external conditions in the labs.

In Fig. \ref{fig:24hours}, we plot the SKR over 24 hours with all 7 co-propagating classical channels on. The 7 classical channels are not affected by co-propagation with the pulsed local oscillator of the CV-QKD system. Although we do not specifically measure the bit-error-rate, we can conclude this from the eye diagram after the link for all classical channels, which are shown in Fig.~\ref{fig:expSetup}(b). We observe no apparent difference in the eye diagrams when we turn on or  off the CV-QKD system. The average distilled SKR is approximately in the range of 20 to 50  kbit/s over the 24 hours. The SKR is changing over time and we attribute these fluctuations to external conditions, such as temperature change in the room. This claim is supported by the fact that during the night when no one is entering the lab, the SKR is very stable.

\vspace{-5pt}
\section{Conclusions}\vspace{-5pt}
We have demonstrated co-propagation of CV-QKD and 7 classical 12.5~Gbit/s OOK signals over 10~km of fiber. We have demonstrated continuous secret key generation between 20 and 50 kbit/s over 24 hours. These results shows that CV-QKD is a good contestant for co-integration with classical channels in future quantum enabled networks. 

\vspace{7pt}
{\noindent  \footnotesize \emph{This work was partly funded by ImPACT Program of Council for Science, Technology and Innovation (Cabinet Office, Government of Japan). }
}

\end{document}